\documentclass[journal,comsoc,twoside]{IEEEtran}
\usepackage[numbers,sort&compress]{natbib}
\IEEEoverridecommandlockouts

\ifCLASSINFOpdf
\else
\fi
\usepackage[T1]{fontenc}
\usepackage{algorithm}
\usepackage{algorithmicx}
\usepackage{algpseudocode}
\usepackage{multirow}
\usepackage{graphicx}
\usepackage[utf8]{inputenc}

\usepackage{amsmath}
\usepackage{amsfonts}
\begin{document}

%
\title{The Application of Zig-Zag Sampler in Sequential Markov Chain Monte Carlo}
%
%
%

\author{Yu~Han,
        Kazuyuki~Nakamura
\thanks{The authors are with the Graduate School of Advanced Mathematical Sciences, Meiji University, Tokyo,
164-8525, Japan (e-mail: cs201004@meiji.ac.jp; knaka@meiji.ac.jp).}
\thanks{This work has been submitted to the IEEE for possible publication. Copyright may be transferred without notice, after which this version may no longer be accessible.}
\thanks{}}

\maketitle

\begin{abstract}
Particle filtering methods are widely applied in sequential state estimation within nonlinear non-Gaussian state space model. However, the
traditional particle filtering methods suffer the weight degeneracy in the high-dimensional state space model.
Currently, there are many methods to improve the performance of particle filtering in high-dimensional state space model. Among these,
the more advanced method is to construct the Sequential Markov chain Monte Carlo (SMCMC) framework by implementing the
Composite Metropolis-Hasting (MH) Kernel. In this paper, we proposed to discretize the Zig-Zag Sampler and apply
the Zig-Zag Sampler in the refinement stage of the Composite MH Kernel within the SMCMC framework which is implemented
the invertible particle flow in the joint draw stage. We evaluate the performance of proposed method through numerical
experiments of the challenging complex high-dimensional ﬁltering examples. Numerical experiments show that in high-dimensional
state estimation examples, the proposed method improves estimation accuracy and increases the acceptance ratio compared
with state-of-the-art ﬁltering methods.
\end{abstract}


%
\IEEEpeerreviewmaketitle

\section{Introduction}
%
%
%
%
\IEEEPARstart{E}{stimating} signal from a series of noisy observations is an crucial assignment in many applications,
especially in the  weather fields and multi-target tracking \cite{martinez2009bayesian} \cite{creal2015high} \cite{van2015nonlinear}.
Nonlinear non-Gaussian state space model is widely
used in these applications and Sequential Monte Carlo (SMC) \cite{speekenbrink2016tutorial} methods are employed to deal with
the inference problem of state space model.
However, SMC are inefficient in the high-dimensional state sapce model because SMC use the importance sampling method as the bias
correction. Importance sampling could lead to high variance of estimation result. Therefore, in the high-dimensional
situation, the weight of large number of particles is almost zero and this phenomenon is called weight degeneracy. 
Although the variance could be minimized by using optimal importance distribution, it is difficult
to sample from optimal importance distribution \cite{speekenbrink2016tutorial}. As a further matter, the performance of SMC also depends on the design of importance
distribution which need to be designed according to different model \cite{arulampalam2002tutorial}. This undoubtedly increases the difficulty of solving
the filtering problem by implementing the SMC methods.

In the past few decades, several methods for improving the performance of particle filtering have been developed. 
The first kind of methods are auxiliary particle filtering (APF) \cite{pitt1999filtering}, unscented particle ﬁltering \cite{van2000unscented}, and RaoBlackwellised
particle ﬁltering \cite{sarkka2007rao}. These methods aim to approximate the optimal importance distribution by designing proposal distribution.
APF methods improve the sampling efficiency by way of introducing new auxiliary variable. Unscented particle ﬁltering methods use
the unscented transformation to approximate the optimal importance distribution. RaoBlackwellised particle ﬁltering
methods combine Kalman filter \cite{grewal2020kalman} and particle filtering to improve the performance. These methods can only perform well
in some specific situations and state estimation in the high-dimensional situation is still a challenging mission.
The second kind of methods are extended Kalman ﬁlter (EKF) \cite{wan2001dual}, unscented Kalman ﬁlter (UKF) \cite{wan2001unscented}, and the ensemble Kalman ﬁlter (EnKF) \cite{evensen2003ensemble}.
These methods rely on local linearization medium and crude functional approximation method. EnKF and EKF can perform well with the state space model which
have Gaussian error and UKF can perform well with the state space model which have non-Gaussian error. These methods also cannot be completely applied
in the nonlinear non-Gaussian state space model. The third kind of methods are multiple particle ﬁltering \cite{djuric2007multiple}, block particle ﬁltering \cite{doucet2009tutorial}, and
space-time particle ﬁltering \cite{beskos2017stable}. Their core approach is decomposition or partition of state space and transforming a
high-dimensional problem into a low-dimensional problem. But these methods cannot completely solve the problem of “dimension curse”
caused by weight degeneracy.

Compared with SMC, Sequential Markov Chain Monte Carlo (SMCMC) \cite{lamberti2017gradient} are another methods to solve the filtering problem and successfully applied
within the nonlinear non-Gaussian high-dimensional state space model. The main idea of SMCMC is to construct a Markov
chain which stationary distribution is empirical approximation of posterior distribution and sample particles from
the transition kernel. SMCMC combines the recursive nature of SMC and the high performance of Markov Chain Monte Carlo
(MCMC) methods. Besides, SMCMC uses the rejection-acception sampling mechanism as the bias correction. Therefore, the first advantage of SMCMC
is that we do not care about the weight degeneracy in the SMCMC framework. The performance of SMCMC depends on the 
transition kernel. Thus, exploring an efficient transition kernel is an important mission. Luckily, SMCMC could use all
MCMC kernel \cite{andrieu2003introduction} which is another advantage of SMCMC. There are mainly four types of transition kernel of SMCMC in \cite{septier2015langevin}:
Optimal Independent Metropolis-Hasting (MH) Kernel, Approximation of the Optimal Independent MH Kernel, Independent
MH Kernel Based on Prior as Proposal, and Composite MH Kernel. Optimal Independent MH Kernel is difficult to design under normal
circumstances. Compared with Approximation of the Optimal Independent MH Kernel and Independent
MH Kernel Based on Prior as Proposal, Composite MH Kernel performs well in the high-dimensional situation because Composite MH Kernel could
gradually collect information of the target distribution, this means, Composite MH Kernel could explore the neighborhood of the current value of Markov
chain \cite{septier2015langevin}.

Although Composite MH Kernel is more efficient than other SMCMC kernels, there are some methods to further improve the performance of
Composite MH Kernel. The Metropolis Adjusted Langevin Algorithm (MALA) \cite{girolami2011riemann} \cite{xifara2014langevin} \cite{livingstone2014information}
and Hamiltonian Monte Carlo (HMC) \cite{betancourt2017geometric} which based on Riemannian
manifold are proposed as the refinement step in the Composite MH kernel. After combining with Composite MH Kernel, MALA and HMC can have better performance in the high-dimensional
state space model. Reference \cite{septier2015langevin} showed that the sequential manifold HMC (SmHMC) method performs better than sequential MALA (SMALA) method.
On the other hand, the SmHMC method also has a limitation that target distribution needs to satisfy the log-concave condition.
Although there are ways to solve this problem, the expected values of negative Hessian and some hyperparameters are still required.

Recently, a novel class of methods called invertible particle flow \cite{li2017sequential} is proposed to implement and construct the first proposal distribution
in the Composite MH Kernel. Invertible particle flow can guide particles from prior to posterior distribution, that means it could
approximate the true posterior density from the handleable prior density through the intermediate density. Therefore, 
invertible particle flow can be used to construct a proposal distribution which closes to the posterior distribution
in the join draw step of Composite MH Kernel. There are two advantages of implementing invertible particle flow in the Composite MH Kernel. This
combination could better close the Optimal Independent MH Kernel and improve the sampling efficient in the joint draw step. Exact Daum-Huang
(EDH) method and local Exact Daum-Huang (LEDH) \cite{li2017particle} method are one of the invertible particle flow methods and
further improves the performance of Composite MH Kernel.

In recently past, research in non-reversible piecewise deterministic MCMC (PD-MCMC) \cite{vanetti2017piecewise} \cite{sherlock2017discrete} field has made great progress.
The key idea
of PD-MCMC is to use the non-reversible piecewise
deterministic Markov process (PDMP) \cite{fearnhead2018piecewise} \cite{bierkens2018piecewise} to sample from the target distribution
which as its invariant distribution. On one hand,
non-reversible PDMP have better mix properties compared to reversible Markov process. Thus, PD-MCMC could yield estimation
values with low variance. On the other hand, PD-MCMC methods are rejection-free methods, this means, every data sampled 
would not be dropped. Therefore, it could explore the target distribution more efficiently, especially in the case of 
high-dimensional and Big-Data. Classical examples of PDMP are Bouncy Particle Sampler (BPS) \cite{bouchard2018bouncy} and Zig-Zag Sampler \cite{bierkens2019zig}.

Even though there are many advantages of PD-MCMC, it cannot be directly used in the SMCMC framework.
First, if we want to construct the PDMP, we need to simulate the inhomogeneous Poisson process. In complicated situation, inhomogeneous Poisson process is
simulated by the Poisson thinning method. Thus, it requires to know the local upper bound of the derivative of the log 
posterior distribution. However, local upper bound is difficult to obtain. Second, the PDMP methods mentioned above are
all based on continuous conditions. However, discrete condition need to be satisfied in the SMCMC framework. Therefore,
Discrete Bouncy Particle Sampler (DBPS) is proposed in \cite{bouchard2018bouncy}. DBPS is discrete version of BPS and can be
implemented without simulating inhomogeneous Poisson process. Reference \cite{pal2018sequential} proposed that implementing DBPS as refinement step in
the Composite MH Kernel of SMCMC framework and DBPS could further improve the acceptance ratio compared with other MCMC methods.
But like BPS, DBPS still has a phenomenon called reducible. Adding the velocity refresh step in the DBPS can solve reducible problem.
Reference \cite{bouchard2018bouncy} proposes three speed refresh methods: Full refresh, Ornstein-Uhlenbeck refresh, and Brownian motion on the
unit sphere refresh. However, there is no general method to determine the velocity refresh.
We need to choose different velocity method that is suitable for the
situation. Moreover, improper velocity refresh method will greatly reduce the efficiency of the DBPS \cite{hanyuk}.

In this paper, we propose to discretize the Zig-Zag Sampler and combine it with the SMCMC framework as the
individual refinement step. In addition, we still employ the EDH method to construct the first proposal distribution in the Composite MH Kernel of SMCMC framework.
Compared with the advanced methods in this field, our main contributions include: 1) we exploit the mild ergodic condition
of Zig-Zag Sampler to avoid choosing suitable velocity refresh method in the SMCMC framework; 2) we employ the better mix properties of Zig-Zag Sampler which can provide
higher particle diversity and improve the acceptance ratio comparing with traditional SmHMC and MALA methods in the SMCMC framework;
3) we exhibit the effiency of the proposed method in the complex inference scheme with the high-dimensional state space model.

We organize the rest of the paper as below. In Section II, we perform the problem statement. In Section III, we present a brief
review of SMCMC, Composite MH Kernel, EDH, and LEDH. In Section IV, we describe the proposed method which combining discretized Zig-Zag Sampler and EDH in
the SMCMC framework. In Section V, we describe the numerical experiments and exhibit the results. In Section VI, we outline the contributions
and results.

\section{Problem Statement}
\begin{figure}[t]
  \begin{centering}
      \includegraphics[width=3 in]{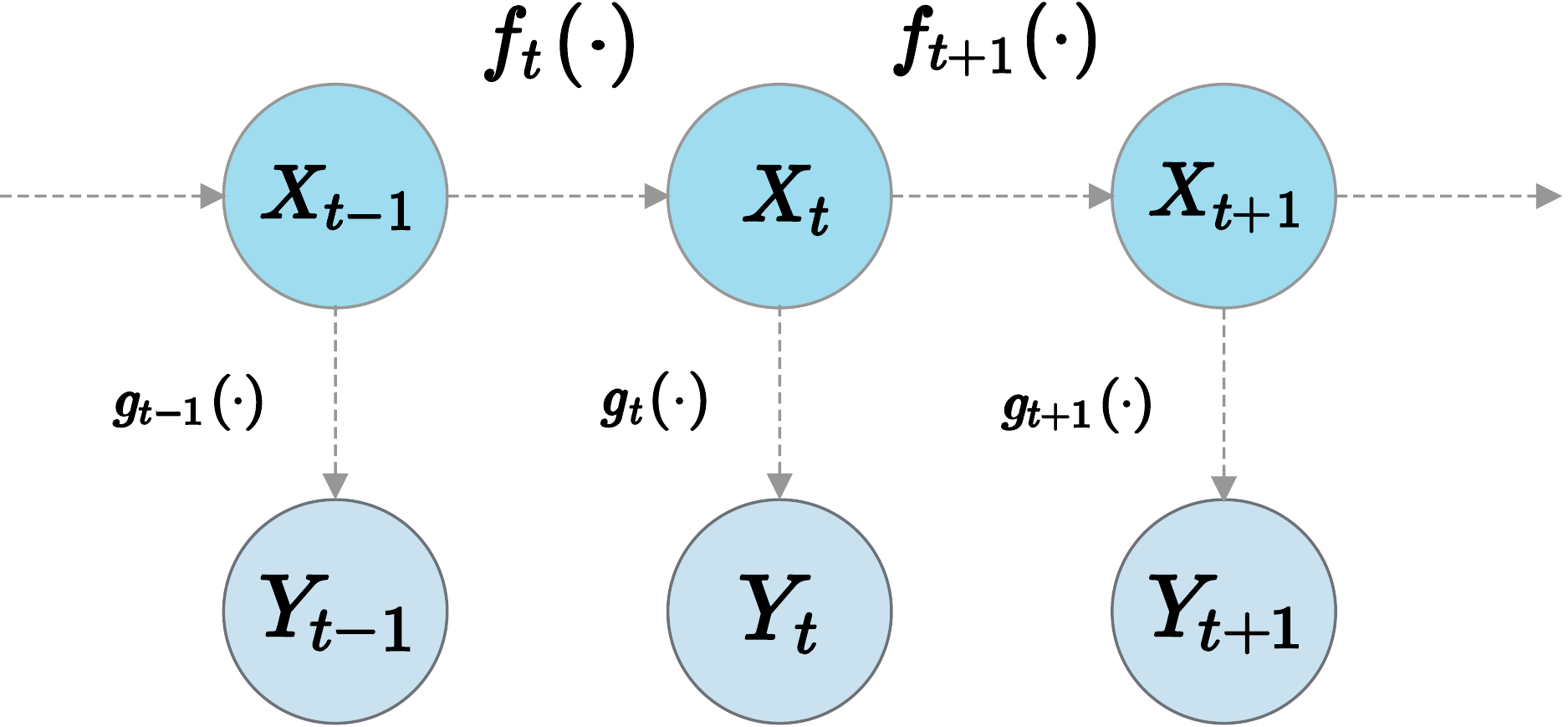}%
      \caption{Graphical representation of state space model.}
  \end{centering}
  \label{fig_graphicmodel_case}
\end{figure}%
In the nonlinear non-Gaussian state space model inference problem, we interested in obtaining the marginal posterior probability density
$p(x_t|y_{1:t})$. $x_t$ is the state of the dynamic system at time $t$ and $y_{1:t}=\{y_1,...,y_t\}$ is a sequence collection of observations up to time
$t$. We can consider to model the state equation and observation equation by the following state sapce model:
\begin{align}
x_{0} &\sim \mu_{0}(x)  \\
x_{t} & =f_{t}\left(x_{t-1}, v_{t}\right) \quad \text { for } t \geq 1,  \\
y_{t} & =g_{t}\left(x_{t}, w_{t}\right) \quad \quad \text { for } t \geq 1.
\end{align}
Here $\mu_0(x)$ is the initial probability density function of the initial state $x_0$
which means the time step before any observations arrives. The function $f_t: \mathbb{R}^{d} \times \mathbb{R}^{d^{\prime}} \rightarrow \mathbb{R}^{d}$
is the state equation which describes the dynamic of the unknown state $x_t \in \mathbb{R}^{d}$ and the system noise $v_t \in \mathbb{R}^{d^{\prime}}$. Observation
$y_t$ is generated by the $g_{t}: \mathbb{R}^{d} \times \mathbb{R}^{C^{\prime}} \rightarrow \mathbb{R}^{C}$
which is the observation equation and describes the relationship between observation $y_t$, state $x_t$,
and the observation noise $w_t \in \mathbb{R}^{d^{\prime}} $. We assume that $f_{t}(\cdot, 0)$ is bounded, and $g_{t}(\cdot, 0)$ is a $C_1$
function, i.e.,$g_{t}(\cdot, 0)$ is differentiable everywhere and its derivatives are continuous.

\section{Background Material}
\subsection{Sequential Markov Chain Monte Carlo Methods}
SMCMC methods are proposed as efficient tool to solve the state estimation task of high-dimensional state space model.
The key idea of SMCMC methods is to implement the MH accept-reject way and use the proposal distribution by which
current states are sampled. Then, after obtaining samples, the target distribution is approximated by them. At time step $t$, we interested in sampling
state from the target distribution:
\begin{align}
\pi_{t}\left(x_{1: t}\right)&=p\left(x_{1: t} \mid y_{1: t}\right) \nonumber \\
&=p\left(x_{1: t-1} \mid y_{1: t-1}\right) \frac{p\left(x_{t} \mid x_{t-1}\right) p\left(y_t \mid x_{t}\right)}{p\left(y_t \mid y_{1: t-1}\right)} \nonumber \\
& \propto p\left(x_{t} \mid x_{t-1}\right) p\left(y_{t} \mid x_{t}\right) p\left(x_{1: t-1} \mid y_{1: t-1}\right) \nonumber \\
& \propto p\left(x_{t} \mid x_{t-1}\right) p\left(y_{t} \mid x_{t}\right) \pi_{t-1}\left(x_{1: t-1}\right)
\end{align}

Unfortunately, it is very difficult to sample from the $\pi_{t-1}\left(x_{1: t-1}\right)$
which is analytically intractable in the general nonlinear non-Gaussian state space model. Therefore, the empirical approximation distribution is proposed to replace
$\pi_{t-1}\left(x_{1: t-1}\right)$. The empirical approximation distribution could be obtained after the previous recursion and defined as:
\begin{align}
\widehat{\pi}_{t-1}\left(dx_{1: t-1}\right)=\frac{1}{N} \sum_{m=N_{b}+1}^{N+N_{b}} \delta_{X_{t-1,1: t-1}^{m}}\left(d x_{1: t-1}\right)
\end{align}
then we replace the distribution of interest with the following equation (\ref{pituta}) \cite{septier2015langevin}:
\begin{align}
\breve{\pi}_{t}\left(x_{1: t}\right) \propto p\left(y_{t} \mid x_{t}\right) p\left(x_{t} \mid x_{t-1}\right) \widehat{\pi}_{t-1}\left(x_{1: t-1}\right) \label{pituta}
\end{align}
Here $N_b$ is the number of burn-in samples and $N$ is the number of reserved samples, $\delta$ is the Dirac delta function.
$\left\{X_{t-1,1: t-1}^{m}\right\}_{m=N_{b}+1}^{N_{b}+N}$ are the $N$ samples obtained from the Markov chain at time $t-1$,
whose stationary distribution is $\breve{\pi}_{t-1}\left(x_{1: {t-1}}\right)$ and $\mathcal{K}_t$ is an MCMC kernel of invariant distribution
$\breve{\pi}_{t}\left(x_{1: t}\right)$. Finally, at the time $t$, we obtain the approximation of marginal posterior distribution by the following equation:
\begin{align}
p\left(x_{t} \mid y_{1: t}\right) \approx \frac{1}{N} \sum_{m=N_{b}+1}^{N+N_{b}} \delta_{X^{m}_{t, 1: t}}\left(d x_{t}\right) \label{app}
\end{align}
In Algorithm 1 \cite{septier2015langevin}, we summarized the general framework of the SMCMC.
\floatname{algorithm}{Algorithm}
\renewcommand{\algorithmicrequire}{\textbf{InPut:}}
\renewcommand{\algorithmicensure}{\textbf{OutPut:}}
    \begin{algorithm}
        \caption{Generic SMCMC framework}
        \begin{algorithmic}[1] 
                \If {time $t = 1$}
                \For{$j=1,...,N+N_b$}
                    \State Sample $X_{1,1}^j \sim \mathcal{K}_1(X^{j-1}_{1,1},\cdot)$ with $\mathcal{K}_1$ an MCMC kernel of invariant distribution $\pi_1(x_1) \propto g_1(y_1 \mid x_1)\mu(x_1) $
                \EndFor
                \ElsIf{time $t \geq 2$}
                \For{$j=1,...,N+N_b$}
                  \State Reﬁne empirical approximation of previous posterior distribution.
                  \State Sample $X_{t,1:t}^j \sim \mathcal{K}_t(X^{j-1}_{t,1:t},\cdot)$ with $\mathcal{K}_1$ an MCMC kernel of invariant distribution $\breve{\pi}_{t}\left(x_{1: t}\right)$ defined in (\ref{pituta})

                \EndFor
                \EndIf
                \Ensure Approximation of the smoothing distribution with the following empirical measure:
                $$\pi\left(x_{1: t}\right) \approx \frac{1}{N} \sum_{j=N_{b}+1}^{N+N_{b}} \delta_{X_{t, 1: t}^{j}}\left(d x_{1: t}\right)$$

        \end{algorithmic}
    \end{algorithm}
\subsection{Composite MH Kernel}
The performance of SMCMC is heavily dependent on MCMC kernel. A properly designed transition kernel can
greatly improve the performance of SMCMC in the high-dimensional situation. Some
different SMCMC kernels are introduced in \cite{septier2015langevin} \cite{septier2009mcmc}. The optimal choice of transition kernel is the Optimal
Independent MH Kernel in which
the proposal distribution is independent of the current value. However, the proposal
distribution is constructed by the target distribution in the framework of Optimal Independent MH Kernel. Therefore, it
is difficult to
adopt this scheme in most cases.

Subsequently, Approximation of the Optimal Independent MH Kernel and Independent MH
Kernel Based on Prior as Proposal are proposed to approximate
the Optimal Independent MH Kernel by using Laplace approximation or local linearization techniques. But it
cannot be easy to approximate accurately and acceptance ratio could be very little in the
complex or high-dimensional situation.

Composite MH Kernel uses local proposal instead of global proposal, so it will reduce the
dependence on the target distribution, and thus be more efficient. The Composite MH Kernel method is summarized in Algorithm 2.
    \begin{algorithm}
        \caption{Composite MH Kernel in a unifying framework of SMCMC \cite{septier2015langevin}}
        \begin{algorithmic}[1] 
                \Require $x^{j-1}_{t,0:t}$
                \Ensure  $x^{j}_{t,0:t}$
                \State $Joint~Draw$
                \State Propose $x_{t, 0: t}^{*} \sim q_{t, 1}(x_{0: t} \mid x_{t, 0: t}^{j-1})$;
                \State Compute the MH acceptance ratio \\
                $\rho_{1}=\min (1, \frac{\breve{\pi}_{t}(x_{t, 0: t}^{*})}{q_{t, 1}(x_{t, 0: t}^{*} \mid x_{t, 0: t}^{j-1})} \frac{q_{t, 1}(x_{t, 0: t}^{j-1} \mid x_{t, 0: t}^{*})}{\breve{\pi}_{t}(x_{t, 0: t}^{j-1})})$;
                \State Accept $x_{t,0:t}^j = x^{*}_{t,0:t}$ with probability $\rho_{1}$, otherwise set $x_{t,0:t}^j = x^{j-1}_{t,0:t}$;
                \State $Individuate~refinement~of~x_{t,0:t-1}^j$;
                \State Propose $x_{t, 0: t-1}^{*} \sim q_{t, 2}(x_{0: t-1} \mid x_{t, 0: t}^{j})$;
                \State Compute the MH acceptance ratio \\
                $\rho_{2}=\min (1, \frac{\breve{\pi}_{t}(x_{t, 0: t-1}^{*}, x_{t, t}^{j})}{q_{t, 2}(x_{t, 0: t-1}^{*} \mid x_{t, 0: t}^{j})} \frac{q_{t, 2}(x_{t, 0: t-1}^{j} \mid x_{t, 0: t-1}^{*}, x_{t, t}^{j})}{\breve{\pi}_{t}(x_{t, 0: t}^{j})})$;
                \State Accept $x_{t,0:t-1}^j = x^{*}_{t,0:t-1}$ with probability $\rho_{2}$, otherwise set $x_{t,0:t-1}^j = x^{j-1}_{t,0:t-1}$;
                \State $Individual~refinement~of~x^j_{t,t}$
                \State Propose $x_{t, t}^{*} \sim q_{t, 3}(x_{t} \mid x_{t, 0: t}^{j})$
                \State Compute the MH acceptance ratio \\
                $\rho_{3}=\min (1, \frac{\breve{\pi}_{t}(x_{t, 0: t-1}^{j}, x_{t, t}^{*})}{q_{t, 3}(x_{t, t}^{*} \mid x_{t, 0: t}^{j})} \frac{q_{t, 3}(x_{t, t}^{j} \mid x_{t, 0: t-1}^{j}, x_{t, t}^{*})}{\breve{\pi}_{t}(x_{t, 0: t}^{j})})$;
                \State Accept $x_{t,t}^j = x^{*}_{t,t}$ with probability $\rho_{3}$, otherwise set $x_{t,t}^j = x^{j-1}_{t,t}$;

        \end{algorithmic}
    \end{algorithm}
Reference \cite{septier2015langevin} proposed to use Composite MH Kernel which composed of joint draw step and refinement step.
It performs more efficient in the high-dimensional situation. Here, $q_{t,1}(\cdot)$ is the first proposal distribution, $q_{t,2}(\cdot)$ is the second proposal distribution, and $q_{t,3}(\cdot)$ is the third proposal distribution.
All path of states are updated by using MH sampler in the joint draw step. In the refinement step,
previous obtained $x_{1:t-1}$ and current $x_t$ are updated in order. In addition, any MCMC kernel could be employed in the Composite MH Kernel.
Therefore, Independent MH Kernel Based on Prior as Proposal usually is implemented in the joint draw step as the first proposal distribution and $\widehat{\pi}_{t-1}$
is implemented in the individual refinement step as the second proposal distribution $q_{t,2}(\cdot)$.
However, the block MH-within Gibbs sampling is employed as the refinement step in
the traditional Composite MH Kernel and it will result in extremely low sampling rate.
In order to improve the efficient, HMC and DBPS \cite{pal2018sequential} is proposed as the refinement
step to construct $q_{t,3}(\cdot)$ in the Composite MH Kernel and these methods are become one class of the most effective methods.
\subsection{Invertible Particle Flow}
Recently, a new class of Monte Carlo-based ﬁltering called invertible
particle flow \cite{daum2008particle} is proposed and perform well in the high-dimensional
state space model. Assume that $N$ particles
$\{x^j_{t-1}\}^N_{j=1}$ are obtained at time step $t-1$. $\{x^j_{t-1}\}^N_{j=1}$ are employed to approximate the target distribution. After implanting by the state equation at time $t$, we obtain
$\{\tilde{x}_{t}^{j}\}_{j=1}^{N}$ which means the
predictive target distribution at time step $t$. Then, particle flow
could migrate the particles and lead the particles approximate the
posterior distribution well.

Invertible particle flow could be implemented as a background stochastic process
$\eta_{\lambda}$ in a pseudo time interval $\lambda \in [0,1]$. Under normal circumstances, we assume that
$\{\tilde{x}_{t}^{j}\}_{j=1}^{N}$ is replaced by $\{\eta_0^j\}_{j=1}^N$. After the whole dynamic stochastic process is completed,
we obtain $\{\eta_1^j\}_{j=1}^N$ which is regarded as the particles that approximate the target distribution.

In \cite{daum2007nonlinear} \cite{daum2009gradient}, the invertible particle flow $\{\eta_0^j\}_{j=1}^N$, follows equation (\ref{ode}):
\begin{align}
    \frac{d \eta_{\lambda}^{j}}{d \lambda}=\zeta(\eta_{\lambda}^{j}, \lambda) \label{ode}
\end{align}
where $\zeta$ is constrained by the Fokker-Planck equation with zero diffusion:
\begin{align}
    \frac{\partial p(\eta_{\lambda}^{j}, \lambda)}{\partial \lambda}= & -p(\eta_{\lambda}^{j}, \lambda) \operatorname{div}(\zeta(\eta_{\lambda}^{j}, \lambda)) \nonumber \\
    & -\frac{\partial p(\eta_{\lambda}^{j}, \lambda)}{\partial \eta_{\lambda}^{j}} \zeta(\eta_{\lambda}^{j}, \lambda) \label{pflow}
\end{align}
where $p$ is the probability density of $\eta_{\lambda}$. In this paper, we mainly introduced two invertible particle flow methods:
Exact Daum and Huang Filter (EDH) method and local Exact Daum and Huang Filter (LEDH) method.
\subsubsection{The Exact Daum and Huang Filter with SMCMC}
The particles flow trajectory of $\eta_{\lambda}$ in the exact Daum and Huang Filter \cite{daum2010exact} \cite{daum2010exactnon} follows this equation:
\begin{align}
    \zeta(\eta_{\lambda}^{j}, \lambda)=A(\lambda) \eta_{\lambda}^{j}+b(\lambda)
\end{align}
Here, $A(\lambda)$ and $b(\lambda)$ is presented in \cite{ding2012implementation}:
\begin{align}
    A(\lambda)&= -\frac{1}{2} P H(\lambda)^{T}\left(\lambda H(\lambda) P H(\lambda)^{T}+R\right)^{-1} H(\lambda) \nonumber \\
    b(\lambda)&= (I+2 \lambda A(\lambda))\left[(I+\lambda A(\lambda))PH(\lambda)^{T} R^{-1}q(\lambda)\right]\nonumber \\
    q(\lambda)&= (y-e(\lambda))+A(\lambda) \bar{\eta}_{0}\nonumber \\
    e(\lambda)&=h\left(\bar{\eta}_{\lambda}, 0\right)-H(\lambda) \bar{\eta}_{\lambda} \nonumber
\end{align}

Here $P$ is the covariance matrix of system noise error in the state equation, $y$
is still the observation, $H$ is the observation matrix and $R$ is the covariance matrix
of the observation noise. For nonlinear observation equation, $H$ requires to
be constructed by the linearization technique which is employed at the mean of
particles $\eta_{\lambda}$ and following the equation (\ref{lin}):
\begin{align}
    H(\lambda)=\left.\frac{\partial h(\eta, 0)}{\partial \eta}\right|_{\eta=\bar{\eta}_{\lambda}} \label{lin}
\end{align}
Numerical integration is employed to approximate the solution of equation (\ref{ode})
when implementing the exact flow algorithm.
Assume that we obtain a set which includes the different time step point. These time step point are $\lambda_0,\lambda_1,..,\lambda_{N_{\lambda}}$
and they satisfy the condition which is $0=\lambda_{0}<\lambda_{1}<\ldots<\lambda_{N_{\lambda}}=1$.
The time interval $\varepsilon_m = \lambda_m-\lambda_{m-1}$
and the sum of total time interval is $1$, i.g., $\sum_{i=1}^{N_{\lambda}}\varepsilon_i=1$. Therefore, the integral between two adjacent of EDH follows this equation:
\begin{align}
    \eta_{\lambda_{m}}^{j} & =f_{\lambda_{m}}(\eta_{\lambda_{m-1}}^{j}) \nonumber \\
    & =\eta_{\lambda_{m-1}}^{j}+\epsilon_{m}(A(\lambda_{m}) \eta_{\lambda_{m-1}}^{j}+b(\lambda_{m})) \nonumber
\end{align}

Thus, by using the Euler update rule \cite{li2019invertible},
\begin{align}
\begin{split}
\eta_{1}^{j} &=f_{\lambda_{N_{\lambda}} }\left(f_{\lambda_{N-1}}\left(\ldots f_{\lambda_{1}}\left(\eta_{0}^{j}\right)\right)\right.\\
&=\left(I+\epsilon_{N_{\lambda}} A\left(\lambda_{N_{\lambda}}\right)\right) \eta_{\lambda_{N_{\lambda}-1}}^{j}+\epsilon_{N_{\lambda}} b\left(\lambda_{N_{\lambda}}\right) \\
&=C \eta_{0}^{j}+D
\end{split}
\end{align}
where,
\begin{align}
    C&=\prod_{m=1}^{N_{\lambda}}\left(I+\epsilon_{N_{\lambda}+1-m} A\left(\lambda_{N_{\lambda}+1-m}\right)\right) \nonumber \\
D&= \epsilon_{N_{\lambda}} b(\lambda_{N_{\lambda}}) \nonumber \\
&+\sum_{l=1}^{N_{\lambda}-1}([\prod_{m=1}^{N_{\lambda}-l}(I+\epsilon_{N_{\lambda}+1-m} A(\lambda_{N_{\lambda}+1-m}))] \epsilon_{m} b(\lambda_{m}))\nonumber
\end{align}

Reference \cite{li2019invertible} proposes to employ invertible particle flow methods to construct the first proposal distribution in the Composite
MH Kernel and gave two conclusions: 1) the mapping involving
$\eta_0^i$ and $\eta_1^i$ is invertible mapping; 2)
$C$ is invertible and the determinant of $C$ is not zero.
Thus, the first proposal distribution $q_{t,1}(\cdot)$ could be constructed by
invertible particle flow mapping and follows equation (\ref{proposal}):
\begin{align}
    q(\eta_{1}^{j} \mid x_{t-1}^{j}, z_{t})=\frac{p(\eta_{0}^{j} \mid x_{t-1}^{j})}{|\operatorname{det}(C)|}  \label{proposal}
\end{align}

The algorithm of applying EDH in Composite MH Kernel is summarized in Algorithm 4.
In the Algorithm 4, we do not need to update $C$ and $D$ at every iteration and
we only need to calculate once before every time step iteration. We draw the
particles from the empirical distribution which is obtained from the previous time iteration. Then,
after calculating the mean of particles, we calculate the auxiliary state $\overline{\eta}_0$ and
employ the $\overline{\eta}_0$ in the $T$ function. Finally, we obtain the parameters
$C$, $D$ and apply them in the SMCMC iterations with every particle at current time.
About $T$ function, it is summarized in the Algorithm 3.
For applying EDH to construct the first proposal distribution in the Composite MH Kernel, the first acceptance ratio is calculated by equation (\ref{accp}):

\begin{small}
\begin{align}
\rho_{1}=\min (1, \frac{p(x_{t, t}^{*(j)} \mid x_{t, t-1}^{*(j)}) p(z_{t} \mid x_{t, t}^{*(j)}) p(\eta_{0}^{j-1} \mid x_{t, t-1}^{j-1})}{p(\eta_{0}^{*(j)} \mid x_{t, t-1}^{*(j)}) p(x_{t, t}^{j-1} \mid x_{t, t-1}^{j-1}) p(z_{t} \mid x_{t, t}^{j-1})}) \label{accp}
\end{align}
\end{small}

\begin{algorithm}
        \caption{$T$ Function: $T(\overline{\eta})=(C,D)$}
        \begin{algorithmic}[1] 
                \Require $\overline{\eta}$
                \Ensure $C~D$
                \State Initialize $C=I,D=0$
                \For{$m=1,...,N_{\lambda}$}
                    \State Set $\lambda_m=\lambda_{m-1}+\varepsilon_m$;
                    \State Caculation $A(\lambda_m) b(\lambda_m)$
                    \State Migrate: $\overline{\eta}:\overline{\eta}=\overline{\eta}+\varepsilon_m(A(\lambda_m)\overline{\eta}+b(\lambda_m))$
                    \State Set $C = (I+\varepsilon_mA(\lambda_m))C$;
                    \State Set $D = (I+\varepsilon_mA(\lambda_m))D+\varepsilon_j(\lambda_m)$;
                \EndFor
        \end{algorithmic}
    \end{algorithm}

\begin{algorithm}
        \caption{Composite MH Kernel with the EDH Particle Flow \cite{li2019invertible}}
        \begin{algorithmic}[1] 
                \Require $x_{t,0:t}^{j-1},\eta^{j-1}_0,C,D$
                \Ensure $x_{t,0:t}^{j},\eta^{j}_0$
                \State $Joint~Draw~of~x_{t,0:t}^{j}$
                \State Draw $x_{t, 0: t-1}^{*(j)} \sim \widehat{\pi}_{t-1}\left(x_{0: t-1}\right)$
                \State Sample $\eta_{0}^{*(j)}=g_{t}(x_{t, t-1}^{*(j)}, v_{t})$
                \State Calculate $x_{t, t}^{*(j)}=C \eta_{0}^{*(j)}+D$
                \State Compute the first MH acceptance ratio $\rho_{1}=$
                $$\min (1, \frac{p(x_{t, t}^{*(j)} \mid x_{t, t-1}^{*(j)}) p(z_{t} \mid x_{t, t}^{*(j)}) p(\eta_{0}^{j-1} \mid x_{t, t-1}^{j-1})}{p(\eta_{0}^{*(j)} \mid x_{t, t-1}^{*(j)}) p(x_{t, t}^{j-1} \mid x_{t, t-1}^{j-1}) p(z_{t} \mid x_{t, t}^{j-1})}) $$
                \State Accept $x^j_{t,0:t}=x_{t, 0: t-1}^{*(j)},\eta^j_0=\eta_{0}^{*(j)}$ otherwise set $x^j_{t,0:t}=x_{t, 0: t}^{j-1},\eta^j_0=\eta_{0}^{j-1}$
                \State $Individuate~refinement~of~x_{t,0:t-1}^j$;
                \State Propose $x_{t, 0: t-1}^{*} \sim \widehat{\pi}_{t-1}(x_{0: t-1})$;
                \State Compute the second MH acceptance ratio $\rho_{2}=$
                $$\min (1, \frac{\breve{\pi}_{t}(x_{t, 0: t-1}^{*}, x_{t, t}^{j})}{q_{t, 2}(x_{t, 0: t-1}^{*} \mid x_{t, 0: t}^{j})} \frac{q_{t, 2}(x_{t, 0: t-1}^{j} \mid x_{t, 0: t-1}^{*}, x_{t, t}^{j})}{\breve{\pi}_{t}(x_{t, 0: t}^{j})})$$
                \State Accept $x_{t,0:t-1}^j = x^{*}_{t,0:t-1}$ with probability $\rho_{2}$, otherwise set $x_{t,0:t-1}^j = x^{j-1}_{t,0:t-1}$;
                \State $Individual~refinement~of~x^j_{t,t}$
                \State Propose $x_{t, t}^{*} \sim q_{t, 3}(x_{t} \mid x_{t, 0: t}^{j})$
                \State Compute the third MH acceptance ratio $\rho_{3}=$
                $$\min (1, \frac{\breve{\pi}_{t}(x_{t, 0: t-1}^{j}, x_{t, t}^{*})}{q_{t, 3}(x_{t, t}^{*} \mid x_{t, 0: t}^{j})} \frac{q_{t, 3}(x_{t, t}^{j} \mid x_{t, 0: t-1}^{j}, x_{t, t}^{*})}{\breve{\pi}_{t}(x_{t, 0: t}^{j})})$$
                \State Accept $x_{t,t}^j = x^{*}_{t,t}$ with probability $\rho_{3}$, otherwise set $x_{t,t}^j = x^{j-1}_{t,t}$;
                \State Calculate $\eta^j_0=C^{-1}(x_{t,t}^j-D)$;

        \end{algorithmic}
    \end{algorithm}
\subsubsection{The Local Exact Daum and Huang Filter with SMCMC}
Comparing with the EDH, the system is linearized and the drift
term is updated for each individual particle in the LEDH \cite{li2017particle}. Therefore, the trajectory of every particles
follows:
\begin{align}
    \zeta\left(\eta_{\lambda}^{j}, \lambda\right)=A^{j}(\lambda) \eta_{\lambda}^{j}+b^{j}(\lambda)
\end{align}
where
\begin{align}
    A^{j}(\lambda)&=-\frac{1}{2} P H^{j}(\lambda)^{T}\left(\lambda H^{j}(\lambda) P H^{j}(\lambda)^{T}+R\right)^{-1} H^{j}(\lambda) \nonumber\\
    b^{j}(\lambda)&=\left(I+2 \lambda A^{j}(\lambda)\right)\left[\left(I+\lambda A^{j}(\lambda)\right) P H^{j}(\lambda)^{T}\right. \nonumber \\
    &\left.\times R^{-1}\left(y-e^{j}(\lambda)\right)+A^{j}(\lambda) \bar{\eta}_{0}\right] \nonumber
\end{align}
The integral between two adjacent of LEDH follows this equation:
\begin{align}
\eta_{\lambda_{m}}^{j} &=f_{\lambda_{m}}^{j}(\eta_{\lambda_{m-1}}^{j})  \nonumber\\
&=\eta_{\lambda_{m-1}}^{j}+\epsilon_{m}(A^{j}(\lambda_{m}) \eta_{\lambda_{m-1}}^{j}+b^{j}(\lambda_{m}))
\end{align}
The algorithm of combining LEDH with Composite MH Kernel is summarized in Algorithm 5.
\section{Proposal Method}
The Zig-Zag Sampler \cite{bierkens2019zig} \cite{bierkens2019large} in the continuous time situation can be seen as a discrete set of velocity \cite{bierkens2017limit}.
When the state space is the $d$-dimensional, from a physical point of view,
the velocity $v$ follows:
\begin{align}
    v=\sum_{i=1}^dv_i\textbf{e}_i \nonumber %
\end{align}
Here $\{\textbf{e}_1, \textbf{e}_2,…, \textbf{e}_d\}$ is a set of orthongal
base vectors in the state space of $\mathbb{R}^d$. From a probabilistic point of view, the Zig-Zag Sampler can be considered as constructing by $d$-distinct different event
type which have their own different event rate and simulate their different Possion process independently.
We denote that the event rate of type $i$
is $\lambda_i(\textbf{x},\textbf{v})$ and thus the general event rate formulation of the Zig-Zag Sampler is
\begin{align}
    \lambda(\textbf{x},\textbf{v})=\sum_{i=1}^d\lambda_i(\textbf{x},\textbf{v})
\end{align}
Therefore,
after simulating the next event time in every direction by using $\lambda_i(\textbf{x},\textbf{v})$ in equation (\ref{inp}),
\begin{align}
    \mathbb{P}(\tau_{i} \geq t)=\exp (-\int_{0}^{t}\lambda_{i}(x(s), v) ds) \label{inp}
\end{align}
we pick the
smallest event time $\tau_{i_0}$ with the direction as the next event time
\begin{align}
    i_{0}:=\operatorname{argmin}_{i \in\{1, \ldots, d\}} \tau_{i} \nonumber
\end{align}
and flip the velocity in this direction with the
probability $\lambda_i(\textbf{x},\textbf{v})/\lambda(\textbf{x},\textbf{v})$.
We denote that the flip function is $F_i[v]$. It means that flipping the $i$-th dimension of $v$, and follows:
$$
\left(F_{i}[v]\right)_{j}=\left\{\begin{array}{ll}
v_{j} & \text { if } i \neq j \\
-v_{j} & \text { if } i=j
\end{array}\right.
$$
In the most case, one of the important
property of Zig-Zag Sampler is that the event rate could construct the relationship with the invariant
distribution $\Phi(x)$ and it satisfies the following equation (\ref{lad}):
\begin{align}
    \lambda_i(x)={max(0,v_i\partial_i\Phi(x))}_+ \label{lad}
\end{align}
Therefore, we can use equation (\ref{lad}) to simulate the Possion process by target distribution in the every direction.

However, we cannot directly apply traditional Zig-Zag Sampler in the SMCMC framework because it needs satisfy the discrete condition.
Thus, inspired by the Zig-Zag Sampler method and in order to be able to be applied in the SMCMC framework,
we used the idea of DBPS for reference and proposed to discretize the Zig-Zag Sampler. We call it the discretized Zig-Zag Sampler (DZZ).

Comparing with the troditional Zig-Zag Sampler, in the DZZ method, when the sample is rejected,
we can think the event has happend at a certain direction. Thus, we calculate the dot product with every direction, like in the Zig-Zag Sampler. Becasue we want to
avoid to simulate the inhomogenous Possion process, we directly obtain these dot product values as the weight of $d$-distint directions. The greater the weight, the more likely it is that the event will occur in this direction.
Based on this set of weight, we randomly sample a direction index in all directions as the direction to flip the $i$-th dimension velocity $v_i$ of the velocity $v$.
$v_i$ is the component of $v$ in that direction.
Thus, we can implement this idea as the DZZ to construct the transition kernel.

\begin{algorithm}[t]
        \caption{Composite MH Kernel with the LEDH Particle Flow \cite{li2019invertible}}
        \begin{algorithmic}[1] 
                \Require $x_{t,0:t}^{j-1},\eta^{j-1}_0,C^{j-1},D^{j-1}$
                \Ensure $x_{t,0:t}^{j},\eta^{j}_0,C^{j},D^{j}$
                \State $Joint~Draw~of~x_{t,0:t}^{j}$
                \State Draw $x_{t, 0: t-1}^{*(j)} \sim \widehat{\pi}_{t-1}(x_{0: t-1})$
                \State Sample $\eta_{0}^{*(j)}=g_{t}(x_{t, t-1}^{*(j)}, v_{t})$
                \State Calculate $\overline{\eta}_{0}^{*(j)}=g_{t}(x_{t, t-1}^{*(j)}, 0)$
                \State Perform $T$ function
                $$
                T(\overline{\eta}_{0}^{*(j)}) = (C^{*(j)},D^{*(j)})
                $$
                \State Calculate $x_{t, t}^{*(j)}=C^{*(j)} \eta_{0}^{*(j)}+D^{*(j)}$
                \State Compute the first MH acceptancce ratio $\rho_{1}=$\\
                $\min (1, \frac{p(x_{t, t}^{*(j)} \mid x_{t, t-1}^{*(j)}) p(z_{t} \mid x_{t, t}^{*(j)})|\operatorname{det}(C^{*(j)})| p(\eta_{0}^{j-1} \mid x_{t, t-1}^{j-1})}{p(\eta_{0}^{*(j)} \mid x_{t, t-1}^{*(j)}) p(x_{t, t}^{j-1} \mid x_{t, t-1}^{j-1}) p(z_{t} \mid x_{t, t}^{j-1})|\operatorname{det}(C^{j-1})|})$;
                \State Accept $x^j_{t,0:t}=x_{t, 0: t-1}^{*(j)},\eta^j_0=\eta_{0}^{*(j)},C^j=C^{*(j)},D^j=D^{*(j)}$ otherwise set $x^j_{t,0:t}=x_{t, 0: t}^{j-1},\eta^j_0=\eta_{0}^{j-1},C^j=C^{j-1},D^j=D^{j-1}$
                \State $Individuate~refinement~of~x_{t,0:t-1}^j$;
                \State Propose $x_{t, 0: t-1}^{*} \sim \widehat{\pi}_{t-1}(x_{0: t-1})$;
                \State Compute the second MH acceptance ratio \\
                $\rho_{2}\!=\!\min (1, \frac{\breve{\pi}_{t}(x_{t, 0: t-1}^{*}, x_{t, t}^{j})}{q_{t, 2}(x_{t, 0: t-1}^{*} \mid x_{t, 0: t}^{j})} \frac{q_{t, 2}(x_{t, 0: t-1}^{j} \mid x_{t, 0: t-1}^{*}, x_{t, t}^{j})}{\breve{\pi}_{t}(x_{t, 0: t}^{j})})$;
                \State Accept $x_{t,0:t-1}^j = x^{*}_{t,0:t-1}$ with probability $\rho_{2}$, otherwise set $x_{t,0:t-1}^j = x^{j-1}_{t,0:t-1}$;
                \State $Individual~refinement~of~x^j_{t,t}$
                \State Propose $x_{t, t}^{*} \sim q_{t, 3}(x_{t} \mid x_{t, 0: t}^{j})$
                \State Compute the third MH acceptance ratio \\
                $\rho_{3}=\min (1, \frac{\breve{\pi}_{t}(x_{t, 0: t-1}^{j}, x_{t, t}^{*})}{q_{t, 3}(x_{t, t}^{*} \mid x_{t, 0: t}^{j})} \frac{q_{t, 3}(x_{t, t}^{j} \mid x_{t, 0: t-1}^{j}, x_{t, t}^{*})}{\breve{\pi}_{t}(x_{t, 0: t}^{j})})$;
                \State Accept $x_{t,t}^j = x^{*}_{t,t}$ with probability $\rho_{3}$, otherwise set $x_{t,t}^j = x^{j-1}_{t,t}$;
                \State Calculate $\eta^j_0=(C^j)^{-1}(x_{t,t}^j-D^j)$;

        \end{algorithmic}
    \end{algorithm}

The Metropolis within Gibbs method is implemented to sample a candidate state value from proposal
distribution in the individual refinement step of Composite MH Kernel. We proposed to use the DZZ algorithm to replace
the Metropolis within Gibbs method and constrcut the the third proposal distribution $q_{t,3}(\cdot)$ of $x_{t,t}^j$
in Composite MH Kernel. Liking the DBPS algorithm \cite{pal2018sequential}, the DZZ algorithm is implemented by $N_{thinning}$
iterations. At the beginning, we set the initialized recursion value $x_t^{(0)} = x_{t,t}^j$ and finally
we obtain the $x_t^{(N_{thinning})}$ as the refinement value of $x_{t,t}^j$ after $N_{thinning}$ iterations.
Pseudocode is summarized in Algorithm 6.

The delayed rejection and random walk are employed in the Algorithm 6.
In the every iteration of DZZ, auxiliary state is sampled from the
joint distribution $p{(x_t,v)}$ in the two dimensional space and the interested posterior
distribution $\Phi(x_t)$ is the marginal distribution of $p{(x_t,v)}$. The velocity $v$ is used to
explore the state space as an auxiliary variable by regional movement. Its marginal has a spherical symmetric feature.

Before implementing $j$-th iteration of the DZZ, the preconditioning matrix $M$ needs to be obtained and it could improve
efficiency of auxiliary velocity.
Especially, the target distribution has obviously different scales
in various dimensions. After multiplying the $M^{(-1)}$, it is not necessary that finding the more appropriate size of step to decide next
regional movement location. At the same time, it cannot affect the better performance of the DZZ method.
Like in \cite{pal2018sequential}, We employ $\Gamma$ to obtain the $M$. $\Gamma$ follows the equation (\ref{gammmma})
\begin{align}
    \Gamma \approx-\left.\mathbb{E}_{z_{t} \mid x_{t}}\left[\nabla^{2} \ln \phi\left(x_{t}\right)\right]\right|_{x_{t}=x_{t}^{(0)}} \label{gammmma}
\end{align}
here, $\phi\left(x_{t}\right)$ is the target distribution and follows the equation (\ref{pphi})
\begin{align}
    \phi\left(x_{t}\right) &= \breve{\pi}(x_{t, 0: t-1}^{j}, x_{t}) \nonumber \\
    & \propto p(x_{t} \mid x_{t, t-1}^{j}) p\left(z_{t} \mid x_{t}\right) \label{pphi}
\end{align}

Next, at the beginning of each iteration,
a movement is firstly proposed from the previous location and velocity $(x_{t}^{(i-1)}, v^{(i-1)})$ to the current location and velocity $(x_{t}^{(i-1)}+v^{(i-1)},-v^{(i-1)})$. It is reversible with the
location and velocity. MH acceptance is employed to decide whether
accept this sample. Regardless of whether the move is accepted or
rejected, the flip movement of the direction with $v$ is always followed.
Finally, the two reversible movements construct an non-reversible
Markov process, in which state variable moves in the same direction until a rejection occurs.

If the first proposal of location and velocity is rejected, we need to calculate the gradient at the rejection point
and then calculate the $m(t)$ by equation (\ref{m}):
\begin{align}
    m(t)=<v^{\prime(i)},\nabla\ln\phi(x_{t}^{\prime(i)})> \label{m}
\end{align}
Here, $< >$ is the dot product and
the every dimension of $m(t)$ is $m_{[k]}(t), k = 1,\cdots, d$, which is the dot
product between $\nabla \ln \phi(x_{t}^{\prime(i)})_{[k]}$ and $v^{\prime(i)}_{[k]}$.
$\nabla \ln \phi(x_{t}^{\prime(i)})_{[k]}$ is the $k$-th dimension of the gradient and $v^{\prime(i)}_{[k]}$ is the $k$-th dimenison of the velocity $v^{\prime(i)}$.

After obtaining the $m(t)$, we sample a number $\mathcal{I}$ from the set $W = \{1, 2, \cdots, d\}$ which including all dimensions. $max(0,m(t))$ as the sample weight of each dimension.
$\mathcal{I}$  is used to decide the dimension of flipping velocity.
Thus, we flip the $\mathcal{I}$-th dimension of the $v^{\prime(i)}$ and let it as the $v^{\prime\prime(i)}$.
After the delayed rejection, the acceptance probability of the second proposal maintains a detailed balance with the target distribution.
Finally, we obtain iteration value
$x^{(N_{thinning})}_{t}$ as the state value $x_{t,t}^j$. This is the implementation process of the DZZ.

The method of combining EDH and DZZ in the Composite MH Kernel of SMCMC framework is called DZZ (EDH) and the method of
combining LEDH and DZZ in the Composite MH Kernel of SMCMC framework is called DZZ (LEDH).
\begin{algorithm}
        \caption{Individual Refinement step by implementing DZZ }
        \begin{algorithmic}[1] 
                \Require $x_{t,t}^{j},N_{thinning}$
                \Ensure $refinement~of~x_{t,t}^{j}$
                \State $x_t^{(0)} = x_{t,t}^j$
                \State Compute $\Gamma $
                \State Apply the Cholesky decomposition of $\Gamma$ to calculate Compute matrix $M$: $M = \Gamma\Gamma^{T}$
                \State Sample the initial auxiliary velocity $v_{aux}^{(0)}$ from the uniform distribution
                \State Compute $v^{(0)} = M^{-1}v_{aux}^{(0)}$
                \For{$i=1~to~N_{thinning}$}
                \State First Propose $$(x_{t}^{\prime(i)}, v^{\prime(i)})=(x_{t}^{(i-1)}+v^{(i-1)},-v^{(i-1)})$$
                \State Compute the first acceptance ratio
                       $$\rho_1(x_{t}^{(i-1)}, x_{t}^{\prime(i)})=\min (1, \frac{\phi(x_{t}^{\prime(i)})}{\phi(x_{t}^{(i-1)})})$$
                \State Accept the first proposal and go to the Update step, if reject, go to the second proposal step.
                \State Second Propose $(x_{t}^{\prime\prime(i)}, v^{\prime\prime(i)})$ where
                $$v^{\prime\prime(i)} = (F_{\mathcal{I}}[v^{\prime(i)}])_k$$
                $$x_{t}^{\prime \prime(i)}=x_{t}^{(i-1)}+v^{(i-1)}-v^{\prime \prime(i)}$$
                \State Compute second acceptance ratio
                $$\rho_2 = \min (1, \frac{(1-\rho_1(x_{t}^{\prime \prime(i)}, x_{t}^{\prime(i)}))}{(1-\rho_1(x_{t}^{(i-1)}, x_{t}^{\prime(i)}))} \frac{\phi(x_{t}^{\prime \prime(i)})}{\phi(x_{t}^{(i-1)})})$$
                \State Accept the second proposal, otherwise set $(x_{t}^{(i)}, v^{(i)}) \leftarrow (x_{t}^{(i-1)}, v^{(i-1)})$
                \State Update $(x_{t}^{(i)}, v^{(i)}) \leftarrow(x_{t}^{(i)},-v^{(i)})$
                \EndFor
                \State $x_{t,t}^j = x_t^{(N_{thinning})}$
        \end{algorithmic}
\end{algorithm}
\section{Numerical Experiments and Results}
In this section, we conduct two numerical simulation experiments to evaluate the performance of the proposed method.
The ﬁrst numerical experiment is the simplest special model of a dynamic process, dynamic Gaussian process with Gaussian likelihood.
The second numerical experiment is the high-dimensional inference problem in which the state equation is the multivariate
Generalized Hyperbolic (GH) skewed-$t$ distribution and the observation equation is the Poisson distribution. Especially, we contrast
the efficiency of the proposed method and the optimal Kalman ﬁlter in the first numerical simulation experiment.
\subsection{Dynamic Gaussian Process With Gaussian Observations}
\subsubsection{Simulation Setup}

\begin{table}[t]
  \centering
  \caption{MSE, Acceptance Ratio, and execution time per step (Average over 120 trials) in the linear Gaussian example of Large spatial sensor networks example with different observation noise level and dimension}
  \setlength{\tabcolsep}{1mm}{
    \begin{tabular}{|c|c|c|c|c|c|c|c|c|}
    \hline
    \multirow{2}[0]{*}{$\sigma^2_y$} & \multirow{2}[0]{*}{$d$} & \multirow{2}[0]{*}{Algorithm} & \multirow{2}[0]{*}{Particles} & \multirow{2}[0]{*}{MSE} & \multicolumn{3}{c|}{Acceptance Ratio} & \multirow{2}[0]{*}{Time(s)} \\
    \cline{6-8}
          &       &       &       &       & $\rho_1$  & $\rho_2$  & $\rho_3$  &  \\
    \hline
    \multirow{14}[0]{*}{1} & \multirow{7}[0]{*}{64} & SmHMC & 200   & 0.22  & 0.0033 & 0.0064 & 0.88  & 0.07 \\
    \cline{3-9}
          &       & SmHMC(EDH) & 200   & 0.21  & 0.0024 & 0.0088 & 0.88  & 0.11 \\
          \cline{3-9}
          &       & DBPS(EDH) & 30000 & 0.2   & 0.0001 & 0.003 & 0.86  & 22 \\
          \cline{3-9}
          &       & DZZ(EDH) & 30000 & 0.19  & 0.0001 & 0.0001 & 0.96  & 23 \\
          \cline{3-9}
          &       & DZZ(LEDH) & 5000  & 0.2   & 0.0001 & 0.0013 & 0.96  & 9.7 \\
          \cline{3-9}
          &       & PF    & 10000 & 0.5   & NA    & NA    & NA    & 0.09 \\
          \cline{3-9}
          &       & KF    & NA    & 0.1   & NA    & NA    & NA    & 0.001 \\
          \cline{2-9}
          & \multirow{7}[0]{*}{144} & SmHMC & 200   & 0.21  & 0.0029 & 0.0077 & 0.81  & 0.21 \\
          \cline{3-9}
          &       & SmHMC(EDH) & 200   & 0.2   & 0.0024 & 0.0095 & 0.81  & 0.35 \\
          \cline{3-9}
          &       & DBPS(EDH) & 30000 & 0.23  & 0.0001 & 0.0021 & 0.89  & 73 \\
          \cline{3-9}
          &       & DZZ(EDH) & 30000 & 0.19  & 0.0001 & 0.0001 & 0.99  & 72 \\
          \cline{3-9}
          &       & DZZ(LEDH) & 5000  & 0.2   & 0.0001 & 0.0012 & 0.99  & 51 \\
          \cline{3-9}
          &       & PF    & 10000 & 1.05  & NA    & NA    & NA    & 0.19 \\
          \cline{3-9}
          &       & KF    & NA    & 0.17  & NA    & NA    & NA    & 0.0023 \\
          \hline
          \hline
    \multirow{14}[0]{*}{2} & \multirow{7}[0]{*}{64} & SmHMC & 200   & 0.39  & 0.0038 & 0.0066 & 0.87  & 0.07 \\
    \cline{3-9}
          &       & SmHMC(EDH) & 200   & 0.39  & 0.0023 & 0.0086 & 0.88  & 0.11 \\
          \cline{3-9}
          &       & DBPS(EDH) & 30000 & 0.37  & 0.0001 & 0.0032 & 0.86  & 28 \\
          \cline{3-9}
          &       & DZZ(EDH) & 30000 & 0.36  & 0.0001 & 0.0001 & 0.92  & 28 \\
          \cline{3-9}
          &       & DZZ(LEDH) & 5000  & 4.38  & 0.001 & 0.0019 & 0.92  & 9.6 \\
          \cline{3-9}
          &       & PF    & 10000 & 0.59  & NA    & NA    & NA    & 0.08 \\
          \cline{3-9}
          &       & KF    & NA    & 0.32  & NA    & NA    & NA    & 0.0018 \\
          \cline{2-9}
          & \multirow{7}[0]{*}{144} & SmHMC & 200   & 0.35  & 0.0027 & 0.0076 & 0.81  & 0.21 \\
          \cline{3-9}
          &       & SmHMC(EDH) & 200   & 0.35  & 0.0022 & 0.009 & 0.82  & 0.36 \\
          \cline{3-9}
          &       & DBPS(EDH) & 30000 & 0.38  & 0.0001 & 0.0023 & 0.88  & 74 \\
          \cline{3-9}
          &       & DZZ(EDH) & 30000 & 0.33  & 0.0001 & 0.0001 & 0.97  & 74 \\
          \cline{3-9}
          &       & DZZ(LEDH) & 5000  & 13    & 0.0001 & 0.0019 & 0.97  & 51 \\
          \cline{3-9}
          &       & PF    & 10000 & 1.15  & NA    & NA    & NA    & 0.19 \\
          \cline{3-9}
          &       & KF    & NA    & 0.29  & NA    & NA    & NA    & 0.0024 \\
    \hline
    \end{tabular}}
  \label{tableG}%
\end{table}%

In the first numerical simulation experiment,
we explore the performance of proposed method with simplest examination of linear Gaussian model
which is implemented in the large special sensor networks which is proposed in the \cite{septier2015langevin}.
$d$ sensors are uniformly arranged on the plane of a two-dimensional grid. $\{1,2,...,\sqrt {d}\} \times \{1,2,...,\sqrt {d}\}$ is the size of two-dimensional grid.
Each sensor independently collects noisy observation about the phenomenon of interest in its particular location.
We define $x_t^{i} \in \mathbb{R}$ is the state value of $i$-th sensor at time $t$ and thus
$y_t^{i} \in \mathbb{R}$ is the observation value of $i$-th sensor at time $t$.

State equation and observation equation of this linear Gaussian model are, respectively:
\begin{align}
x_t &= \alpha x_{t-1}+v_t,\\
y_t &= x_t +w_t
\end{align}
where $\alpha=0.9$, and $w_t\in\mathbb{R}^d$ follows the Gauss distribution with the
mean ${\mu}_{w_t}=0$, covariance ${\Sigma}_y=\sigma_y^2\textbf{I}_{d\times d}$. $v_t\in\mathbb{R}^d$
follows the Gauss distribution with the mean ${\mu}_{v_t}=0$, covariance ${\Sigma}$.
The disperson matrix $\Sigma$ is denoted by eqation (\ref{sig}):
\begin{align}
    [\Sigma]_{i j}=\alpha_{0} \exp (-\frac{\left\|\mathcal{S}_{i}-\mathcal{S}_{j}\right\|_{2}^{2}}{\beta})+\alpha_{1} \delta_{i j} \label{sig}
\end{align}
where ${\| \cdot \|}_2$ is the $L_2$ norm, the coordinate of the $i$-th sensor is $\mathcal{S}_i \in \mathbb{R}^2$, $\delta_{i,j}$ is the Kronecker delta symbol function.
This equation means that when the value of spatial distance with two sensors declines, the noise dependence increases.
Like in \cite{septier2015langevin}, we set $\alpha_0=3, \alpha_1=0.01, \beta=20$ and the initial value $x_0^i=0$, for $i=1,2,...,d$.
\subsubsection{Parameter Values for the Filtering Algorithms}
We compared the proposed method with the SmHMC, SmHMC(EDH), DBPS(EDH), Particle filtering (PF), and
Kalman filter (KF). For the SmHMC(EDH), the metric is set by equation (\ref{Gx}):
\begin{align}
    G(x_n)=\Sigma_y^{-1}+\Sigma^{-1} \label{Gx}
\end{align}
We conduct the numerical simualtion with different observation noise level, time step,
and the dimension of state. Because the premise of the Kalman filter is the linear Gaussian model, we set the estimated
result of Kalman filter as the optimal result and standard.
The
parameters of SmHMC follow \cite{septier2015langevin}, the parameters of SmHMC(EDH) follow \cite{li2019invertible} ,and the parameters of DBPS follow \cite{pal2018sequential}.
The target of this numerical simulation experiment is
to compare the performance of proposed method with other methods which could provide the optimal result or
near-optimal performance.
\subsubsection{Experimental Results}

We report the average mean square error (MSE), acceptance ratio (if applicable)  over $120$ simulation trials
and execution time per step in Table \ref{tableG}. Besides, the $\sigma^2_y$ is set to $1$, $2$, respectivly.

For this linear Gaussian model, the Kalman filter achieves the minimum error in state estimation.
Because the linear Gaussian model follows the assumptions of Kalman filter, this result is expected.
Proposed DZZ(EDH) method can perform better than the other methods except Kalman Filter method. At the same time,
DZZ(EDH) could provide the higher acceptance ratio than the other SMCMC methods, which means the combination of non-reversible
Markov process and invertible particle flow implements more initializations, better mixing and particle diversity.

However, we observed that another proposed DZZ(LEDH) method only perform well in the case of $\sigma^2_y=1$.
In the remaining cases, DZZ(LEDH) performs very poorly comparing with other methods. Therefore,
we can obtain this conclusion that implementing particle flow methods with every particle provides additional error in DZZ method
with linear Gaussian model. In addition, particle filtering method suffered weight degeneracy problem and perform also poorly.
By combining the partcile flow, the SmHMC(EDH) performs better than the vanilla SmHMC because particle could provide
more suitable initialization particles. Because DBPS(EDH) is based on the non-reversible Markov process, DBPS(EDH) achieves smaller
MSE than the SmHMC(EDH) which is based on the reversible Markov process in the most case. Besides, a suitable velocity refresh method in the DBPS(EDH)
is difficult to find and inappropriate velocity refresh method can influence the performance of DBPS(EDH). Thus,
DBPS(EDH) performance is not as good as DZZ(EDH).

\subsection{Dynamic Skewed-$t$ Process With Poisson Observations}
\subsubsection{Simulation Setup}
In this section, we explore the performance of proposed method in the high-dimensinal complex numerical simulation experiment.
The dynamic model is the non-linear non-Gaussian model. The state equation (\ref{gh}) follows the multivariate Generalised
hyperbolic distribution and it is a heavy-tailed distribution. It is normally applied in the environmental monitoring, weather forecast, and healthy care fields.
\begin{align}
p\left(x_{t} \mid x_{t-1}\right) \propto K_{\lambda-d / 2}\left(\sqrt{\left(\chi+Q\left(x_{t}\right)\right)\left(\psi+\gamma^{T} \Sigma^{-1} \gamma\right)}\right) \nonumber \\
\times \frac{e^{\left(x_{t}-\mu_{t}\right)^{T} \Sigma^{-1} \gamma}}{\sqrt{\left(\chi+Q\left(x_{t}\right)\right)\left(\psi+\gamma^{T} \Sigma^{-1} \gamma\right)}^{\frac{d}{2}-\lambda}} \label{gh}
\end{align}
The observation equation follows the multivariate Poisson distribution (\ref{Po}) and every observation is independent.
\begin{align}
    p\left(y_{t} \mid x_{t}\right)=\prod_{k=1}^{d} \mathcal{P}_{o}\left(y_{t}(k) ; m_{1} \exp \left(m_{2} x_{t}(k)\right)\right) \label{Po}
\end{align}
Here, $\mu_t=\alpha x_{t-1}$, $\alpha$ is a scalar, $K_{\lambda-d / 2}$ is the second kind modefied Bessel function and the order is $\lambda-d / 2$,
$Q\left(x_{t}\right)=\left(x_{t}-\mu_{t}\right)^{T} \Sigma^{-1}\left(x_{t}-\mu_{t}\right)$, $\Sigma$ is as same as like in equation (\ref{sig}),
$\gamma$ and $\nu$ are shape parameters which determine the shape of GH distribution.
We set $\alpha = 0.9$, $\nu=7$, $\gamma={[0.3,...,0.3]}_{1 \times d}$, $m_1=1$, $m_2=\frac{1}{3}$.
$d$ is set by $144$ and $400$, respectively, to explore the estimation performance of various method in different dimension.
The number of experiment is 120 times and the time step is $10$.
\subsubsection{Parameter Values for the Filtering Algorithms}
In this numerical experiment, proposed method is compared with SmHMC(EDH), DBPS(EDH), particle ﬂow particle ﬁlters (PFPF(EDH), PFPF(LEDH)) \cite{li2017particle}, and different Kalman filter.
SmHMC(EDH) is showed the smallest MSE in the \cite{septier2015langevin} and the DBPS(EDH) exhibts the higher accptance ratio in \cite{pal2018sequential}.
Especially in the SmHMC(EDH) method, the metric follows:
\begin{align}
    G\left(x_{n}\right)=\Lambda\left(x_{n}\right)+\widetilde{\Sigma}^{-1}
\end{align}
where $\Lambda\left(x_{n}\right)$ is a matrix with the diagonal elements:
\begin{align}
    [\Lambda\left(x_{n}\right)]_k = m_1m_2^2exp(m_2x_n(k))
\end{align}
and the covariance is given by
\begin{align}
    \widetilde{\Sigma}&=\operatorname{Var}_{p}\left(X_{t} \mid X_{t, t-1}^{j}\right) \nonumber \\
    &=\frac{\nu}{\nu-2} \Sigma+\frac{\nu^{2}}{(2 \nu-8)\left(\frac{\nu}{2}-1\right)^{2}} \gamma \gamma^{T}
\end{align}
The performance of DZZ(EDH) and DZZ(LEDH) is evaluated by using $10000$ particle numbers and $300$ particle numbers, respectively, because
DZZ(LEDH) is much more computationally demanding.
Because the observation noise is related to state value, we need to update the observation
covariance $R$ in EDH and LEDH part. Besides, in UPF method, we set $2d + 1$ sigma points for every paticles.
Every initial state value is set by $0$ with all dimension of state in every filter method.
Specificlly, in the DZZ(EDH), SmHMC(EDH), PFPF(EDH), and DBPS(EDH) Algorithms, we employ the $\bar{\eta}$ to update the matrix $R$.
Besides, $\bar{\eta}^{i}$ is used with every particles in the DZZ(LEDH) and PFPF(LEDH) methods.
\subsubsection{Experimental Results}
\begin{figure}[t]
    \begin{centering}
        \includegraphics[width=3.5 in]{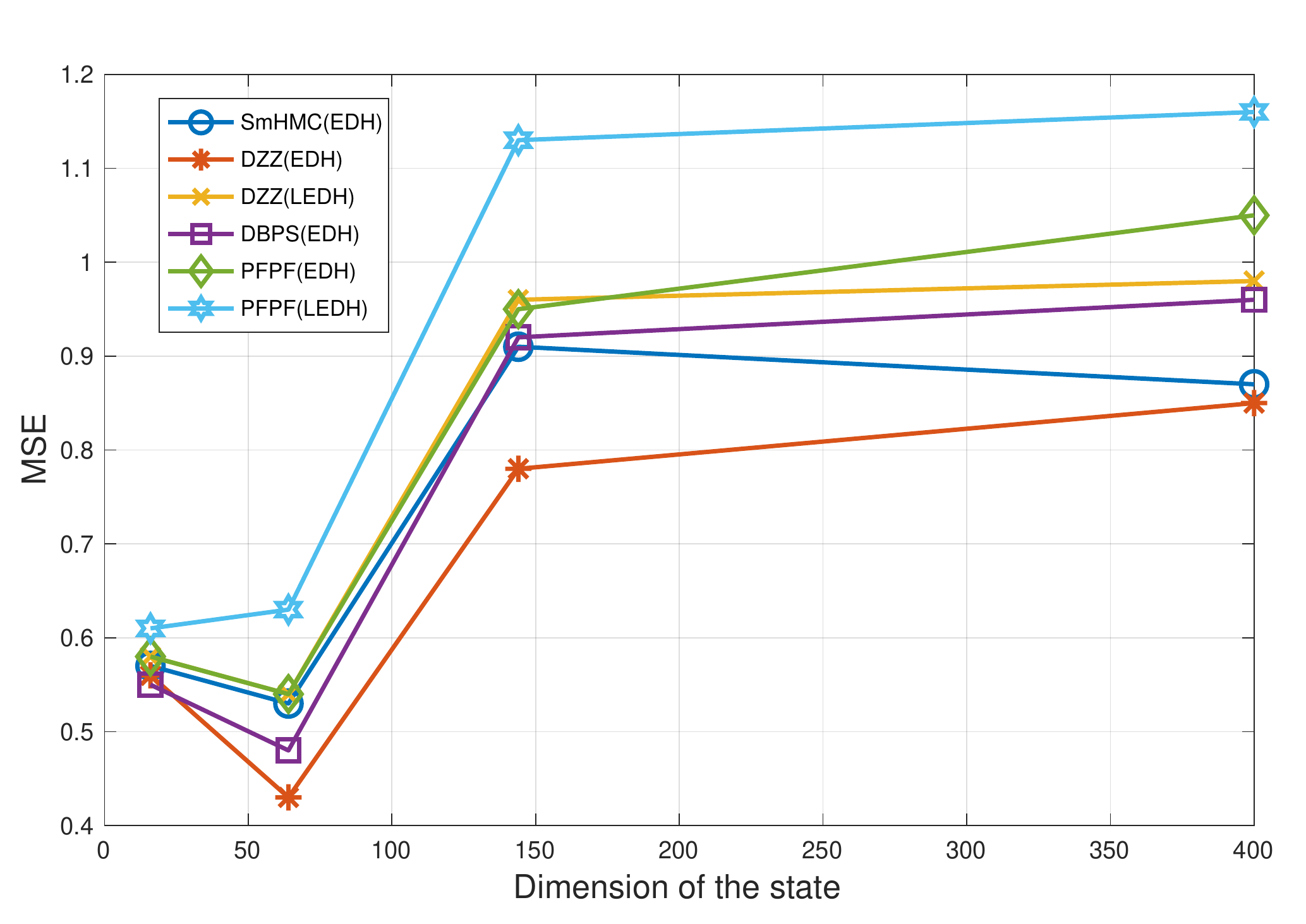}%
        \caption{MSE for the different methods as the dimension of the state increases.}
    \end{centering}
    \label{fig_differentdimension_case}
\end{figure}
\begin{figure}[t]
  \begin{centering}
      \includegraphics[width=3.4 in]{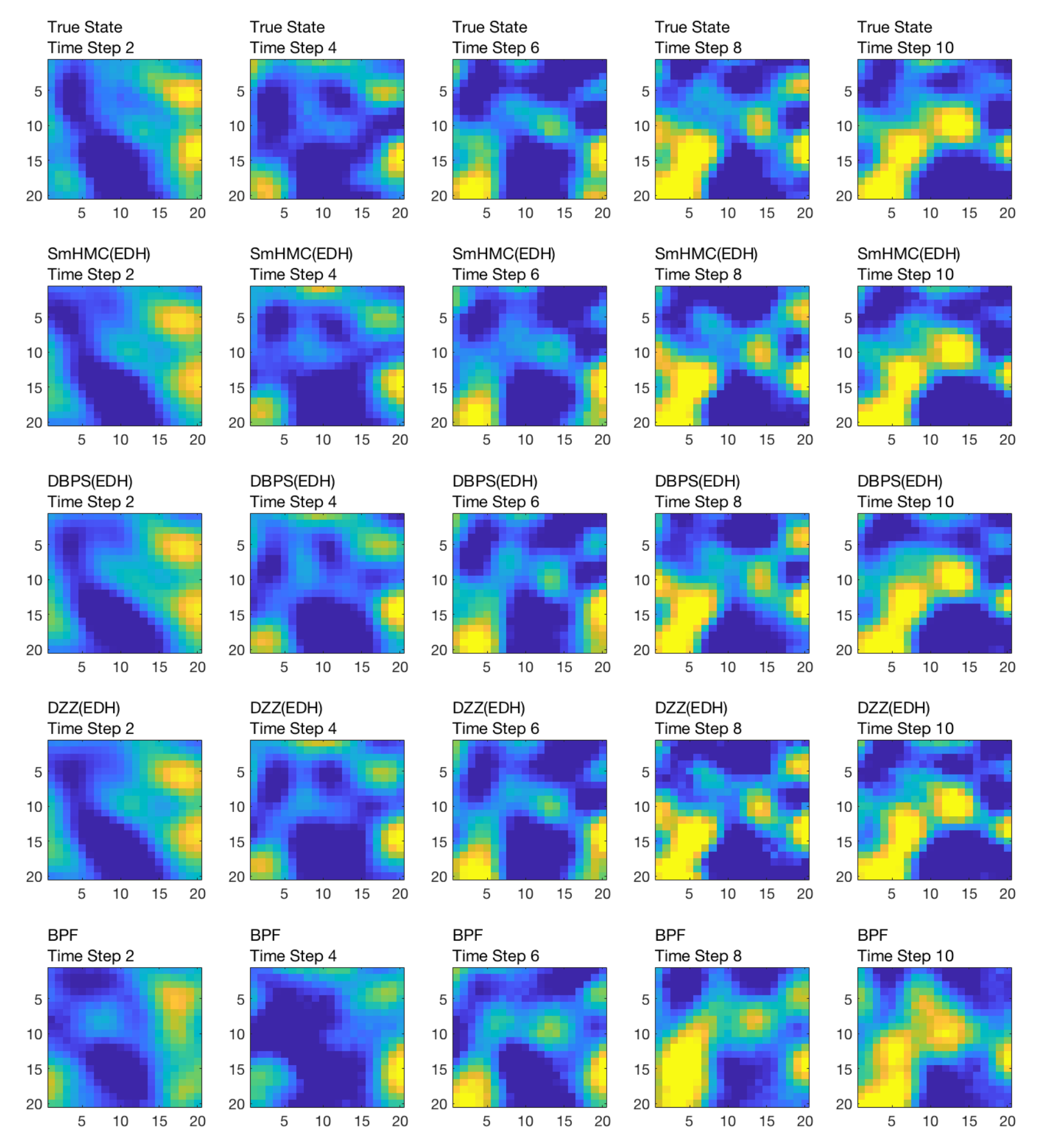}%
      \caption{Illustration of the approximation result at different time steps for several Algorithms ($d = 400$)}
  \end{centering}
  \label{fig_first_case}
\end{figure}

\begin{table}[t]
  \centering
  \caption{MSE, Acceptance Ratio, and execution time per step (Average over 120 trials)
   with Generalised hyperbolic state distribution and Poisson Observations in the Large spatial sensor networks simulation}
  \setlength{\tabcolsep}{1mm}{
    \begin{tabular}{|c|c|c|c|c|c|c|c|}
    \hline
    \multirow{2}[0]{*}{$d$} & \multirow{2}[0]{*}{Algorithm} & \multirow{2}[0]{*}{Particles} & \multirow{2}[0]{*}{MSE} & \multicolumn{3}{c|}{Acceptance} & \multirow{2}[0]{*}{Time(s)} \\
    \cline{5-7}
          &       &       &       & $\rho_1$  & $\rho_2$  & $\rho_3$  &  \\
    \hline
    \multirow{9}[0]{*}{144} & SmHMC(EDH) & 300   & 0.91  & 0.05  & 0.006 & 0.73  & 4.3 \\
          \cline{2-8}
          & DBPS(EDH) & 10000 & 0.92  & 0.02  & 0.001 & 0.91  & 4.6 \\
          \cline{2-8}
          & DZZ(EDH) & 10000 & 0.78  & 0.06  & 0.001 & 0.78  & 5 \\
          \cline{2-8}
          & DZZ(LEDH) & 300   & 0.96  & 0.12  & 0.015 & 0.57  & 4.9 \\
          \cline{2-8}
          & PFPF(EDH) & 10000 & 0.95  & NA    & NA    & NA    & 0.3 \\
          \cline{2-8}
          & PFPF(LEDH) & 300   & 1.13  & NA    & NA    & NA    & 40 \\
          \cline{2-8}
          & BPF   & 100000 & 2.02  & NA    & NA    & NA    & 0.4 \\
          \cline{2-8}
          & UKF   & NA    & 2.2   & NA    & NA    & NA    & 0.0056 \\
          \cline{2-8}
          & EKF   & NA    & 2.52  & NA    & NA    & NA    & 0.0007 \\
    \hline
    \hline
    \multirow{9}[0]{*}{400} & SmHMC(EDH) & 300   & 0.87  & 0.02  & 0.0121 & 0.57  & 60 \\
          \cline{2-8}
          & DBPS(EDH) & 10000 & 0.96  & 0.01  & 0.0001 & 0.93  & 67 \\
          \cline{2-8}
          & DZZ(EDH) & 10000 & 0.85  & 0.03  & 0.0014 & 0.69  & 61 \\
          \cline{2-8}
          & DZZ(LEDH) & 300   & 0.98  & 0.06  & 0.0176 & 0.65  & 120 \\
          \cline{2-8}
          & PFPF(EDH) & 10000 & 1.05  & NA    & NA    & NA    & 2.18 \\
          \cline{2-8}
          & PFPF(LEDH) & 300   & 1.16  & NA    & NA    & NA    & 103 \\
          \cline{2-8}
          & BPF   & 100000 & 2.91  & NA    & NA    & NA    & 1.84 \\
          \cline{2-8}
          & UKF   & NA    & 2.26  & NA    & NA    & NA    & 0.09 \\
          \cline{2-8}
          & EKF   & NA    & 1.96  & NA    & NA    & NA    & 0.02 \\
          \cline{2-8}
    \hline
    \end{tabular}}
  \label{table:skewess}
\end{table}
Fig. \ref{fig_differentdimension_case} shows the MSE of different methods with different dimension. We can find that almost DZZ(EDH) could achieve the smaller MSE with the
different dimension. We visualize the true state and estimation state of different methods by Fig. \ref{fig_first_case}.

Table \ref{table:skewess} detailedly reports the MSE, acceptance ratio, and execution time over $120$ simulation trials.
Since computational burden of different method is completely different from each other, particle numbers are adjusted to balance the accuracy and computation time. Then, we evaluate all methods.
We observed that in the proposed DZZ(EDH) method, the average MSE is reduced compared with
other methods. Comparing with the SmHMC(EDH), because the propeties of non-reversible Markov process and the acceptance
ratio of second refinement step increases which means particles can still keep diversity after the second refinement stage, MSE of DZZ(EDH) is reduced.
Comparing with the DBPS(EDH), the reason of average MSE decreased is that the mild ergodic condition of Zig-Zag method.
It means we do not to set a suitable velocity refresh in the DZZ(EDH) method. DZZ(LEDH) has worse performance than DZZ(EDH).
Therefore, we can indicate a conclusion. It brings additional gain in the DZZ method that we employ independently particle flow methods with every particle.
Therefore, comparing with DZZ(LEDH), the DZZ(EDH) method is adopted in this situation for the reason that it is computationally much more efficient
and acheives smaller average MSE than the DZZ(LEDH) method.

PFPF(EDH) and PFPF(LEDH) perform worse than DZZ(EDH) in the setting with same particle numbers because importance
sampling is employed to close the target posterior distribution in the PFPF methods.
EDH flow and LEDH flow can move particles to a special locality. In this locality, the values of target distribution are comparatively huge.
However, weight degeneracy problem still disturbs the achievement of PFPF(EDH) and PFPF(LEDH) in high-dimensional situation.
The MSE of bootstrap particle filtering (BPF) methods is huge even we use $1$ million particles. Besides, the condition of UKF and EKF
is not be satisfied by this experiment. Therefore, their performance is not good either.

\section{Conclusion}
We have adopt the key idea of Zig-Zag Sampler and implemented DZZ method as the
second refinement step of the Composite MH Kernel which is employed to construct the SMCMC framework. The combination of EDH and
DZZ in SMCMC framework provides the minimal MSE. At the same time, compared with the state-of-the-art methods, proposed method can greatly increase acceptance ratio in the second refinement stage
which means the DZZ(EDH) based SMCMC framework could explore the high-dimensional state space more efficiently.

We evaluate the proposed method in two simulation examples. In the first numerical experiment, DZZ(EDH) method can achieve the smallest MSE. However, DZZ(EDH) needs more computation time
in this setting with large amount of information when the dimension is set by $64$ or $144$ among all particle
filtering methods. At the same time, the proposed DZZ(EDH) method performs the most higher acceptance ratio in the refinement
step which means it can keep the particles diversity. In the second numerical experiment,
when the calculation time is almost the same, the proposed DZZ(EDH) method performs the smallest MSE comparing other SMCMC
and SMC methods. Moreover, the acceptance ratio in the refinement step of DZZ(EDH) is still higher than that of SmHMC(EDH).

In recent years, in addition to Zig-Zag Sampler method, PDMP methods have new developments,
such as Boomerange Sampler method \cite{bierkens2020boomerang}. Therefore, considering the application of the Boomerange Sampler method in the SMCMC
framework is a future important research direction. Besides,
we need to do more comprehensive evaluation with the numerical experiments of DZZ(EDH) method and DZZ(LEDH) method in the future.
These experiments should research the influence of changes
in the initial state value, variance of the state noise and observation noise, and other parameters.
These experiments can further detect the robustness of the algorithm and expose the shortcomings to
promote us to further improve the performance of the algorithm.


%



\section*{Acknowledgment}
This work was supported in part by JST, PRESTO, JPMJPR1774.
The authors thank Y. Liu, M. Coates, S. Pal, F. Septier and G. W. Peters for
publicizing MATLAB Codes associated to \cite{septier2015langevin} \cite{pal2018sequential} \cite{li2019invertible}.
\ifCLASSOPTIONcaptionsoff
  \newpage
\fi



\bibliographystyle{IEEEtran}
\bibliography{IEEEabrv,IEEEabrv1}
\end{document}